\documentclass[aip,preprint]{revtex4-1}

\usepackage{graphicx}
\usepackage{dcolumn}
\usepackage{bm}
\usepackage{amssymb}
\usepackage[T1]{fontenc}
\usepackage{textcomp}
\newcommand{\units}[3][]{$#1\mathrm{#2\,#3}$}
\newcommand{\mum}{\mbox{\textmu m}}

\begin{document}

\title{Spin-wave excitation and propagation in microstructured waveguides of yttrium iron garnet (YIG) /Pt bilayers }

\author{P. Pirro}
\affiliation{Fachbereich Physik and Landesforschungszentrum OPTIMAS, Technische Universit\"at
Kaiserslautern, D-67663 Kaiserslautern, Germany}
\author{T. Br\"acher}
\affiliation{Fachbereich Physik and Landesforschungszentrum OPTIMAS, Technische Universit\"at
Kaiserslautern, D-67663 Kaiserslautern, Germany}
\affiliation{Graduate School Materials Science in Mainz, Gottlieb-Daimler-Strasse 47, D-67663 Kaiserslautern, Germany}
\author{A. Chumak}
\affiliation{Fachbereich Physik and Landesforschungszentrum OPTIMAS, Technische Universit\"at
Kaiserslautern, D-67663 Kaiserslautern, Germany}
\author{B. L\"agel}
\affiliation{Fachbereich Physik and Landesforschungszentrum OPTIMAS, Technische Universit\"at
Kaiserslautern, D-67663 Kaiserslautern, Germany}
\author{C. Dubs}
\affiliation{Innovent e.V., Pr\"ussingstra\ss e 27B, 07745 Jena, Germany}
\author{O. Surzhenko}
\affiliation{Innovent e.V., Pr\"ussingstra\ss e 27B, 07745 Jena, Germany}
\author{P. G\"ornet}
\affiliation{Innovent e.V., Pr\"ussingstra\ss e 27B, 07745 Jena, Germany}
\author{B. Leven}
\affiliation{Fachbereich Physik and Landesforschungszentrum OPTIMAS, Technische Universit\"at
Kaiserslautern, D-67663 Kaiserslautern, Germany}
\author{B. Hillebrands}
\affiliation{Fachbereich Physik and Landesforschungszentrum OPTIMAS, Technische Universit\"at
Kaiserslautern, D-67663 Kaiserslautern, Germany}

\date{\today}

\begin{abstract}
We present an experimental study of spin-wave excitation and propagation in microstructured waveguides patterned from a 100 nm thick yttrium iron garnet (YIG)/platinum (Pt) bilayer. The life time of the spin waves is found to be more than an order of magnitude higher than in comparably sized metallic structures despite the fact that the Pt capping enhances the Gilbert damping. Utilizing microfocus Brillouin light scattering spectroscopy, we reveal the spin-wave mode structure for different excitation frequencies. An exponential spin-wave amplitude decay length of \units{31}{\mum} is observed which is a significant step towards low damping, insulator based micro-magnonics.
\end{abstract}


\maketitle
 
\begin{figure}[h]
\begin{center}
\scalebox{1}{\includegraphics[width=7.0 cm, clip]{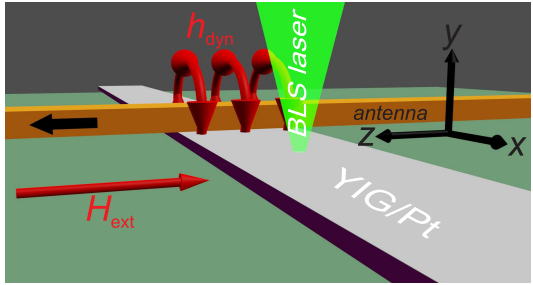}}
\end{center}
\caption{\label{sample} Sample schematic: In a \units{5}{\mum} wide waveguide patterned from a bilayer of YIG/Pt (\units{100}{nm}/\units{9}{nm}), spin waves are excited using the dynamic Oersted fields of a microwave current flowing in a copper antenna. An external bias field $H_\mathrm{ext}$ is applied along the short axis of the waveguide. The spin-wave intensity is detected using microfocus Brillouin light scattering spectroscopy.}
\end{figure}

The concept of magnon spintronics, i.e., the transport and manipulation of pure spin currents in the form of spin-wave quanta, called magnons, has attracted growing interest in the recent years \cite{Serga2010,Jungfleisch2011,Chumak2012,Pirro2011,Lenk2011,Braecher2011,Sandweg2011,Demidov2009,Sebastian2012,Ulrichs2013,Jungfleisch2013,Kelly2013}. One of the key advantages of magnon spin currents is their large decay length which can be several orders of magnitude higher than the spin diffusion length in conventional spintronic devices based on spin-polarized electron currents \cite{}. Considering possible applications, the miniaturization of magnonic circuits is of paramount importance. Up to now, downscaling has been achieved using metallic ferromagnets like NiFe or Heusler compounds \cite{Pirro2011,Lenk2011,Demidov2009,Braecher2011,Sebastian2012,Ulrichs2013}. But even the best metallic ferromagnets exhibit a damping which is two orders of magnitude larger than for Yttrium Iron Garnet (YIG), a ferrimagnetic insulator \cite{Serga2010,SagaofYIG,Glass1976}. However, to the best of our knowledge, as high quality YIG films could only be grown with thicknesses in the range of microns, no microstructured YIG  devices have been fabricated so far. A big step forward has been taken with the recent introduction of methods to produce high quality, low damping YIG films with thicknesses down to several nanometers \cite{Kelly2013,Castel2013,Sun2013,Hahn2013}. In this Letter, we show that microscaled waveguides (see Fig.~\ref{sample}) can be fabricated from liquid phase epitaxy (LPE) grown YIG films of \units{100}{nm} thickness whose high quality has been confirmed by ferromagnetic resonance spectroscopy (FMR). Studying the excitation and propagation of spin-waves in these waveguides by microfocus Brillouin light scattering, we demonstrate that the damping of the unstructured film can be preserved during the structuring process. 

Another key feature of magnon spintronics is its close relationship to a multitude of physical phenomena like spin-pumping, spin-transfer torque, spin Seebeck effect, and (inverse) spin Hall effect, which allow for the amplification, generation and transformation between charge currents and magnonic currents \cite{Tserkovnyak2002,Costache2006,Sandweg2011,Jungfleisch2011,Jungfleisch2013,Sun2013,Heinrich2011,Chumak2012,Hahn2013,Ulrichs2013,Castel2013,Kelly2013,Nakayama2012,Mosendz2010,Qiu2013,Rezende2013}.
Hetero-structures of YIG covered with a thin layer of platinum (Pt) have proven to show these effects which opens a way to a new class of insulator based spintronics. Therefore, we directly study bilayers of YIG/Pt, providing a basis for further studies utilizing the described effects.

The used YIG film is prepared by liquid phase epitaxy from a PbO-B$_2$O$_3$-FeO$_3$ flux melt using a standard isothermal dipping technique with a growth rate of \units{20}{nm/min}. The incorporation of Pb and Pt ions into the garnet lattice allows for a low relative lattice mismatch of $3 \cdot 10^{-4}$.
 
We determine the magnetic properties of the film using FMR and compare the results to measurements performed after the deposition of a \units{9}{nm} Pt film onto YIG using plasma cleaning and RF sputtering.
From the resonance curve $H_\mathrm{FMR}(f_\mathrm{FMR}$), a saturation magnetization of \units[M_\mathrm{s}=]{144 \pm 2}{kA/m} has been determined for the pure YIG film. 
We find that the deposition of Pt slightly reduces the resonance field $\mu_0 H_\mathrm{FMR}$ (for example by \units{1}{mT} for \units[f_\mathrm{FMR}=]{7.0}{GHz}) compared to the pure YIG film. This shift agrees with the recent findings of Ref. \onlinecite{Sun2013}, where a proximity induced ferromagnetic ordering of Pt combined with a static exchange coupling to YIG has been proposed as possible explanation.

\begin{figure}[]
\begin{center}
\scalebox{1}{\includegraphics[width=7.5 cm, clip]{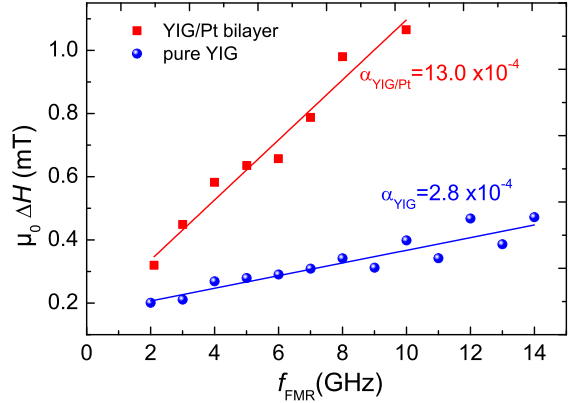}}
\end{center}
\caption{\label{damping} Linewidth $\mu_0 \Delta H$ as a function of the ferromagnetic resonance frequency $f_\mathrm{FMR}$ for the pure YIG film (blue circles) and the YIG/Pt bilayer (red squares). The deviations from the linear increase of $\mu_0 \Delta H$ with $f_\mathrm{FMR}$ (fit according to Eqn.~\ref{delta_H}) are mainly due to parasitic modes causing a small systematical error in the measurement of the linewidth.} 
\end{figure}

Figure~\ref{damping} shows the ferromagnetic resonance linewidth (FWHM) $\mu_0 \Delta H$ with and without Pt and the corresponding fits to evaluate the effective Gilbert damping parameter $\alpha$ according to \cite{Heinrich2011}
\begin{equation}
\label{delta_H}
\mu_0 \Delta H= \mu_0 \Delta H_0+\frac{2 \alpha f_\mathrm{FMR}} {\gamma}
\end{equation}
with the gyromagnetic ratio \units[\gamma=]{28}{GHz/T}. The Gilbert damping $\alpha$ increases by almost a factor of 5 due to the deposition of Pt: from $(2.8\pm0.3)\times 10^{-4}$  to $(13.0\pm1.0) \times 10^{-4}$. The inhomogeneous linewidth $\mu_0 \Delta H_0$ is unchanged within the accuracy of the fit (\units{0.16\pm0.02}{mT} and \units{0.14\pm0.04}{mT}, respectively). Please note that the increase of the damping cannot be explained exclusively by spin pumping from YIG into Pt. Other interface effects, like the already mentioned induced ferromagnetic ordering of Pt in combination with a dynamic exchange coupling may play a role \cite{Sun2013,Rezende2013}.
Using the spin mixing conductance for YIG/Pt (\units[g^{\uparrow \downarrow} \approx]{1.2 \times 10^{18}} {m^{-2}} from \cite{Heinrich2011,Qiu2013}), we find that the expected increase in Gilbert damping due to spin pumping \cite{Nakayama2012,Mosendz2010,Heinrich2011,Jungfleisch2013}
 is \units[\alpha_{sp}=]{1.25\times 10^{-4}}, i.e., it is by a factor of 8 smaller than the measured increase. This clearly demonstrates the importance of additional effects \cite{Sun2013,Rezende2013}. As shown recently in Ref.\cite{Sun2013}, this additional damping can be strongly reduced by the introduction of a thin copper (Cu) layer in between YIG and Pt, which does not significant influence the spin-pumping efficiency.

The micro structuring of the YIG/Pt waveguide is achieved using a negative protective resist mask pattered by electron beam lithography and physical argon ion beam etching. As last production step, a microwave antenna (width \units{3.5}{\mum}, \units{510}{nm} thickness) made of copper is deposited on top of the waveguide (see Fig.~\ref{sample}).

To experimentally detect the spin waves in the microstructured waveguide, we employ microfocus Brillouin light scattering spectroscopy (BLS)\cite{Sebastian2012,Pirro2011,Demidov2009,Jungfleisch2011,Braecher2011,Ulrichs2013}. This method allows us to study the spin-wave intensity as a function of magnetic field and spin-wave frequency. In addition, it  provides a spatial resolution of \units{250}{nm}, which is not available in experiments using spin pumping and inverse spin Hall effect \cite{Chumak2012,Jungfleisch2011,Kelly2013} as these methods integrate over the detection area (and also over the complete spin-wave spectrum\cite{Sandweg2011}).

\begin{figure}[]
\begin{center}
\scalebox{1}{\includegraphics[width=7.5 cm, clip]{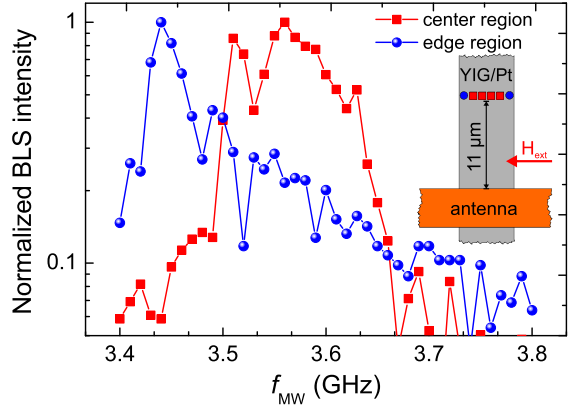}}
\end{center}
\caption{\label{RF_scan}Normalized BLS intensity (log scale) as a function of the applied microwave frequency $f_\mathrm{MW}$ (external field \units[\mu_0 H_\mathrm{ext}=]{70}{mT}). The blue line (circular dots) shows the spectrum measured at the edges of the waveguide (see inset). The red line (rectangular dots) is an average of the spectra recorded in the center of the waveguide.}
\end{figure}

To achieve an efficient spin-wave excitation, we apply a static magnetic field of \units{70}{mT} perpendicular to the long axis of the waveguide. The dynamic Oersted field of a microwave current passing through the antenna exerts a torque on the static magnetization. This configuration results in an efficient excitation of Damon-Eshbach like spin waves which propagate perpendicular to the static magnetization. A microwave power of \units{0}{dBm} (pulsed, duration \units{3}{\mu s}, repetition \units{5}{\mu s}) in the quasi-linear regime, where nonlinearities are not significantly influencing the spin-wave propagation, has been chosen.
To obtain a first characterization of the excitation spectrum, BLS spectra as a function of the applied microwave frequency ($f_\mathrm{MW}$) have been taken at different positions across the width of the waveguide at a distance of \units{11}{\mum} from the antenna. Figure~\ref{RF_scan} shows the spectrum of the edge regions (blue circles) and of the center of the waveguide (red squares, see sketch in the inset). The main excitation in the center of the waveguide takes place at frequencies between \units[f_\mathrm{MW}=]{3.49-3.66}{GHz} and we will refer to these spin-wave modes as the {\it waveguide modes}. At the borders of the waveguide, {\it edge modes} have their resonance around \units[f_\mathrm{MW}=]{3.44}{GHz}. The reason for the appearance of these edge modes is the pronounced reduction of the effective magnetic field $H_\mathrm{eff}$ at the edges by the demagnetization field  and the accompanying  inhomogeneity of the $z$-component of the static magnetization. This situation has been analyzed experimentally and theoretically in detail for metallic systems \cite{Gubbiotti2004,Bayer2004}. 

\begin{figure}[]
\begin{center}
\scalebox{1}{\includegraphics[width=8.5 cm, clip]{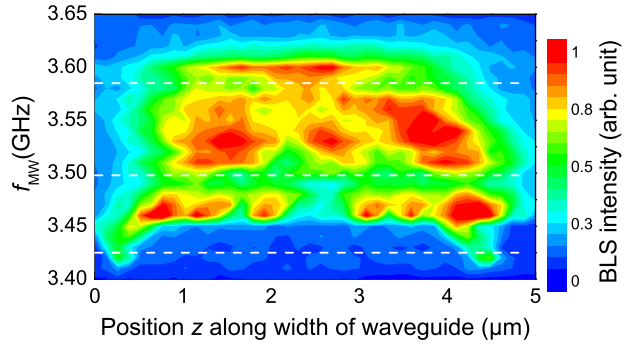}}
\end{center}
\caption{\label{RF_y_scan} BLS intensity (linear scale) as a function of the position along the width of the waveguide for different excitation frequencies $f_\mathrm{MW}$ (\units[\mu H_\mathrm{ext}=]{70}{mT}). Frequencies below \units{3.45}{GHz} show strongly localized edge modes which start to extend into the center of the waveguide for frequencies between \units{3.45-3.50}{GHz}. For higher $f_\mathrm{MW}$, waveguide modes appear which have their local intensity maxima in the center of the waveguide. The dashed lines indicate the calculated minimal frequencies of the waveguide modes shown in Fig.~\ref{dispersion}.}
\end{figure}

To get a better understanding of the nature of the involved spin-wave modes, mode profiles at different excitation frequency measured at a distance of \units{6}{\mum} from the antenna are shown in Fig.~\ref{RF_y_scan}. The evolution of the modes can be seen clearly: for frequencies below \units[f_\mathrm{MW}=]{3.45}{GHz}, the spin-wave intensity is completely confined to the edges of the waveguide. In the range \units[f_\mathrm{MW}=]{3.45-3.50}{GHz}, the maximum of the intensity is also located near the edges, but two additional local maxima closer to the center of the waveguide appear. For frequencies in the range of \units{3.50-3.57}{GHz}, three spin-wave intensity maxima symmetrically centered around the center of the waveguide are observed. This mode is commonly labeled as the third waveguide mode $n=3$ ($n$ denotes the number of maxima across the width of the waveguide). For higher $f_\mathrm{MW}$, only one intensity maximum is found in the center of the waveguide (first waveguide mode, $n=1$).

\begin{figure}[]
\begin{center}
\scalebox{1}{\includegraphics[width=7.5 cm, clip]{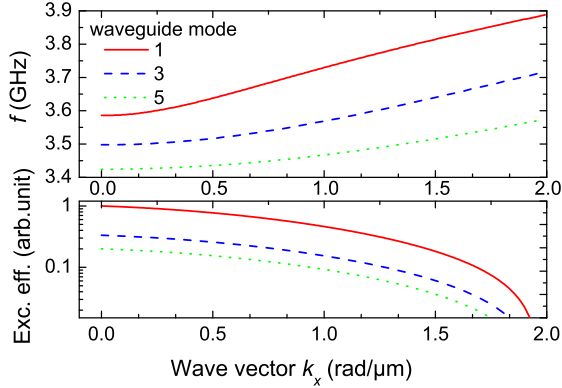}}
\end{center}
\caption{\label{dispersion}(color online) Dispersion relations and amplitude excitation efficiencies for the first three odd waveguide modes of a transversally magnetized YIG waveguide  and an antenna width of \units{3.5}{\mum} (external field \units[\mu H_\mathrm{ext}=]{70}{mT}, width of waveguide \units{5}{\mum}, further parameters see \cite{parameter}).}
\end{figure}


For the waveguide modes, we can compare the experimental results to theoretical considerations. The theory for spin waves in thin films\cite{Kalinikos1986} with the appropriate effective field from micromagnetic simulations and a wave-vector quantization over the waveguide's short axis provides an accurate description of the spin-wave mode dispersions \cite{Pirro2011,Braecher2011,Demidov2009}. Figure~\ref{dispersion} shows the dispersion relations and the excitation efficiencies of the waveguide modes $n=1, 3, 5$. Only odd waveguide modes can be efficiently excited \cite{Demidov2009,Pirro2011} (even modes have no net dynamic magnetic moment averaged over the width of the waveguide) using direct antenna excitation. The minimal frequencies of these three modes are indicated as dashed lines for comparison in Fig.~\ref{RF_y_scan}.
Comparing Fig.~\ref{RF_y_scan} and Fig.~\ref{dispersion}, we find a reasonable agreement between theory and experiment for the first and the third waveguide mode. The $n=5$ and higher waveguide modes are not visible in the experiment. Due to the fact that the excitation efficiency and the group velocity of the spin-wave modes decreases with increasing $n$ (see Ref. \onlinecite{Demidov2009,Pirro2011} for details), this can be attributed to a small amplitude of these modes.

\begin{figure}[]
\begin{center}
\scalebox{1}{\includegraphics[width=8.5 cm,clip]{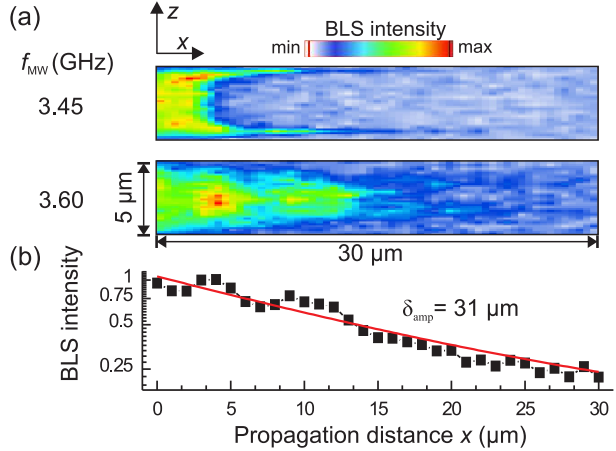}}
\end{center}
\caption{\label{2D_maps}(a) BLS intensity maps (linear scale) for two different excitation frequencies (\units[\mu_0 H_\mathrm{ext}=]{70}{mT}). (b) Integrated BLS intensity (logarithmic scale) over the width of the waveguide for \units[f_\mathrm{MW}=]{3.60}{GHz} including a fit to determine the exponential amplitude decay length (\units[\delta_\mathrm{amp} =]{31}{\mum}).}
\end{figure}

To visualize the influence of the spatial decay of the spin waves on the mode composition, Fig.~\ref{2D_maps}~(a) shows 2D spin-wave intensity maps for two exemplary excitation frequencies. For \units[f_\mathrm{MW}=]{3.45}{GHz}, edge modes can be detected for distances larger than \units{20}{\mum}. Different higher order waveguide modes are also excited, but they can only be detected within \units{5}{\mum} from the antenna. From this findings, we can conclude that the edge modes are dominating the propagation in this frequency range because of their high group velocities (proportional to the decay length) compared to the available waveguide modes ($n \ge 5$).

The situation is completely different for \units[f_\mathrm{MW}=]{3.60}{GHz}. Here, the preferably excited $n=1$ waveguide mode is interfering with the weaker $n=3$ waveguide mode causing a periodic beating effect \cite{Demidov2009,Pirro2011,Sebastian2012} of the measured spin-wave intensity. In this frequency range, no significant contribution of modes confined to the edges is visible.


An important parameter for magnonic circuits and applications is the exponential decay length $\delta_\mathrm{amp}$ of the spin-wave amplitude. To determine $\delta_\mathrm{amp}^\mathrm{exper}$ for \units[f_\mathrm{MW}=]{3.60}{GHz}, we integrate the spin-wave intensity over the width of the waveguide (Fig.~\ref{2D_maps}~(b)) and obtain \units[\delta_\mathrm{amp}^\mathrm{exper}=]{31}{\mum} which is substantially larger than the reported decay lengths in metallic microstructures made of Permalloy or Heusler compounds \cite{Pirro2011,Sebastian2012}. This value can be compared to the expected theoretical value $\delta_\mathrm{amp}^\mathrm{theo}=v_g \tau$ where $v_g$ is the group velocity and $\tau$ is the life time of the spin wave. The Gilbert damping of the unpatterned YIG/Pt bilayer \units[\alpha=]{1.3 \cdot 10^{-3}} measured by FMR corresponds to a life time \units[\tau \approx]{28}{ns} for our experimental parameters. The group velocity $v_g$ can be deduced from the dispersion relations in Fig.~\ref{dispersion} or from dynamic micromagnetic simulations yielding \units[v_{g} \approx]{1.0-1.1}{\mum /ns}, thus \units[\delta_\mathrm{amp}^\mathrm{theo}=]{28-31}{\mum}. The agreement with our experimental findings \units[\delta_\mathrm{amp}^\mathrm{exper}=]{31}{\mum} is excellent, especially if one considers that the plain film values of $\alpha$ and $M_\mathrm{s}$, which might have been changed during the patterning process, have been used for the calculation. This indicates that possible changes of the material properties due to the patterning have only an negligible influence on the decay length of the waveguide modes and that the damping of the spin waves due to the Pt capping is well described by the measured increase of the Gilbert damping.

To conclude, we presented the fabrication of micro-magnonic waveguides based on high quality YIG thin films. Spin-wave excitation and propagation of different modes  in a microstructured YIG/Pt waveguide was demonstrated. As expected, the enhancement of the Gilbert damping due to the Pt deposition leads to a reduced life time of the spin waves compared to the pure YIG case. However, the life time of the spin waves in the YIG/Pt bilayer is still more than an order of magnitude larger than in the usually used microstructured metallic systems. This leads to a high decay length reaching \units[\delta_\mathrm{amp}^\mathrm{exper}]={31}{\mum} for the waveguide modes. One can estimate that the achievable decay length for a similar microstructured YIG/Pt waveguide is \units[\delta_\mathrm{amp}=]{100}{\mum} if a Cu interlayer is introduced to suppress the damping effects which are not related to spin pumping \cite{Sun2013} ($\alpha=\alpha_\mathrm{YIG}+\alpha_{sp}$). Going further, from YIG thin films having the same damping than high quality, micron thick LPE films (\units[\alpha \approx]{4 \times 10^{-5}}, \units[\mu_0 \Delta H \approx ]{0.03}{mT}, Ref. \onlinecite{Glass1976,SagaofYIG}), the macroscopic decay length of \units[\delta_\mathrm{Amp}=]{1}{mm} for micro-magnonic waveguides of pure YIG might be achieved. 

Our studies show that downscaling of YIG preserving its high quality is possible. Thus, the multitude of physical phenomena reported for macroscopic YIG can be transferred to microstructures which is the initial step to insulator based, microscaled spintronic circuits.

\end{document}